\documentclass[prl,twocolumn,groupedaddress]{revtex4}

\usepackage{graphicx}
\usepackage{bm}
\usepackage{braket}
\usepackage{physics}
\usepackage{bbold}
\usepackage{cases}
\usepackage{tabularx}
\usepackage{multirow}
\usepackage{braket}
\usepackage{epstopdf}
\usepackage[urlcolor=blue]{hyperref}
\usepackage{stackengine}

\newcommand{\be}{\begin{equation}}
\newcommand{\ee}{\end{equation}}

\newcommand{\bea}{\begin{eqnarray}}
\newcommand{\eea}{\end{eqnarray}}
\usepackage{color}
\usepackage{dcolumn}
\usepackage{epsfig}
\usepackage{bm}
\usepackage[urlcolor=blue]{hyperref}
\hypersetup{backref, colorlinks=true, linkcolor=blue, citecolor=blue}

\begin{document}

\title{Edelstein effect and supercurrent diode effect}

\author{Noah F. Q. Yuan}
\email{fyuanaa@connect.ust.hk}
\affiliation{Harbin Institute of Technology, Shenzhen, 518055, P. R. China}

\begin{abstract}
We self-consistently calculate the supercurrent diode effect from microscopic models of quasi one- and two-dimensional clean superconductors with spin-orbit coupling under external Zeeman fields, and show that the Edelstein effect is responsible for the supercurrent diode effect. In turn, the supercurrent diode effect may serve as a direct measurement of the Edelstein effect as its application.
\end{abstract}

\maketitle

\textit{\textcolor{blue}{Introduction.}}--- 
Experimentally, nonreicprocal transport in superconducting systems has been found in the fluctuating region \cite{Waka,Yasu,Qin,Hoshi}, where resistance is in general finite and nonlinear transport is significantly enhanced compared to the normal conducting state. Recently, Ando \textit{et. al.} \cite{Ando} realized a superconducting diode that has zero resistance in one direction but finite in the opposite direction, which inspired tremendous discoveries of superconducting diodes in various systems, such as two-dimensional electron gas (2DEG) in quantum wells \cite{Chris}, transition metal dichalcogenides \cite{Banabir,Lorenz}, twisted bilayer graphene \cite{Diez} and twisted trilayer graphene \cite{LinJ}. 
 
Theoretically, as far as we know, superconducting diodes were first proposed by Victor M. Edelstein for three-dimensional (3D) polar superconductors \cite{EdelME,EdelSDE}, based on the Edelstein effect \cite{EdelEE,EdelEE1} where Cooper pair momentum couples to magnetization via spin-orbit coupling (SOC) \cite{EdelEE}. 
In recent theories developed for two-dimensional (2D) superconducting diodes without inversion symmetry \cite{Yuan,Akito,James,Harley,Zhai,Ilic}, 
Cooper pairs can be boosted to finite momentum by external magnetic fields, and as a result, critical currents parallel and anti-parallel to the Cooper pair momentum can be unequal.
Quantitative features of 2D superconducting diodes has been calculated phenomenologically \cite{Yuan,Akito,James} and microscopically \cite{Yuan,Akito,James,Harley,Zhai,Ilic,JXHu}.
However, although phenomenonlogical theories could successfully explain 2D superconducting diodes, microscopic theories for 2D superconducting diodes seem to encounter with obstacles: A systematic calculation by S. Ili\'c and F. S. Bergeret \cite{Ilic} claimed that 2D superconductors with SOC will \textit{not} be superconducting diodes up to the linear order of the magnetic field.

\begin{figure}
\centering
\includegraphics[width=0.4\textwidth]{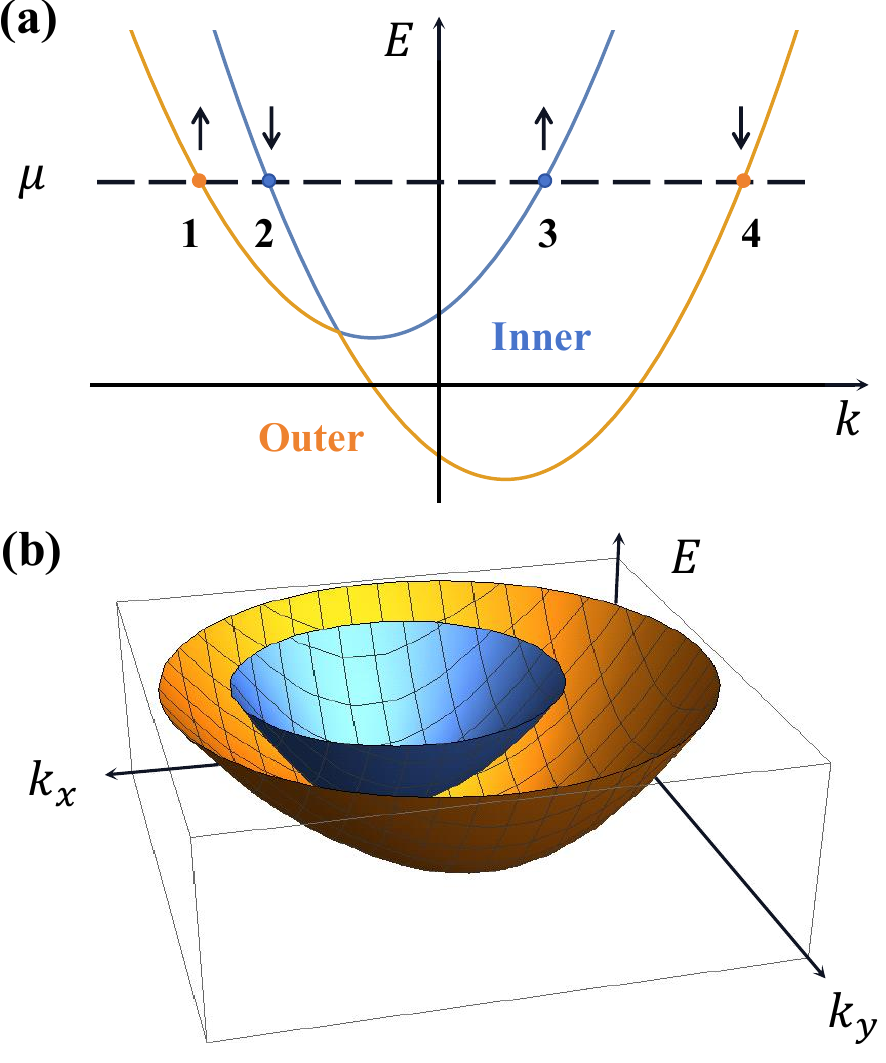}
 \caption{Schematic band structures for (a) one-dimensional wire and (b) two-dimensional electron gas, both with spin-orbit couplings and external Zeeman fields. Different color distinguish energy bands, leading to inner (blue) and outer (orange) Fermi points (a) or Fermi contours (b). Fermi points in (a) are labeled as 1 to 4 with spins indicated by arrows.
  }\label{fig1}
\end{figure}

In this work, we discuss the role of Edelstein effect in 2D superconducting diodes and try to address the issue raised by S. Ili\'c and F. S. Bergeret.
By including the Edelstein effect, 2D superconductors with SOC will become superconducting diodes under external magnetic fields: 
The supercurrent induces magnetization via Edelstein effect, whose polarization direction is pinned by external fields.
Effectively, the external magnetic field favors a particular direction for the supercurrent, which is known as the supercurrent diode effect (SDE), the central topic of this work.

This manuscript is structured as follows. We first consider the one-dimensional (1D) superconductors with SOC to demonstrate the main physics of Edelstein effect in SDE. Then we move to the 2D superconductors with SOC, where previous works \cite{Yuan,Ilic,Akito,James,Harley,Zhai} are also discussed.
Throughout this paper, we work in the clean limit and assume the $s$-wave pairing energy and the Zeeman energy are on similar scales, and Fermi energy is the largest energy scale.

\textit{\textcolor{blue}{1D model.}}---
We consider a generic 1D wire with kinetic energy $\xi$, SOC $g$ and an external Zeeman field $h$
\be\label{eq_1}
H=\xi(k)+g(k)\sigma_z+h\sigma_z
\ee
where $\sigma_z$ is the third Pauli matrix in spin space, and due to time-reversal symmetry, $\xi(-k)=\xi(k)$ is even and $g(-k)=-g(k)$ is odd. The velocity operator 
\be\label{eq_v}
v=\frac{\partial H}{\partial k}=\xi'(k)+g'(k)\sigma_z
\ee
then contains a spin part proportional to the SOC derivative $g'(k)$, which is the orgin of Edelstein effect \cite{EdelEE,EdelEE1}.

We find two bands $E_{\pm}(k)=\xi(k)\pm |g(k)+h|$ in the normal phase.
As superconductivity mainly involves electrons near Fermi energy, in 1D we focus on the four Fermi points where $E_{\pm}=0$, with two inner $(+)$ and two outer $(-)$ Fermi points.
As shown in Fig. \ref{fig1}(a), two bands $E_{\pm}$ are labeled blue $(+)$ and orange $(-)$ respectively, four Fermi points are labeled by 1 to 4 with spins indicated by arrows, where 1 and 4 are outer Fermi points while 2 and 3 are inner. These four Fermi points can form four types of opposite-spin Cooper pairs, denoted as (12), (34), (23) and (14). Among them, Cooper pairs (12) and (34) are formed by electrons from different types of Fermi points, and have nonzero momentum even at zero field, which are usually not favorable at weak fields.
On the other hand, Cooper pairs (23) and (14) are formed by electrons from the same type of Fermi points, and at zero field they become zero-momentum.

At weak fields, we hence focus on Cooper pairs formed by electrons from the same type of Fermi points, whose depairing energy reads
\be\label{eq_D1D}
\mathcal{D}_{\sigma}=E_{\sigma}\left(\frac{1}{2}q+k\right)-E_{\sigma}\left(\frac{1}{2}q-k\right),
\ee
where $\sigma=\pm$ corresponds to Cooper pairs (23) and (14) in Fig. \ref{fig1}(a) respectively.
Corresponding to the velocity operator in Eq. (\ref{eq_v}), the depairing energy 
\be\label{eq_dep}
\mathcal{D}_{\sigma}=\xi'(k)q+\sigma \{g'(k) q+2h\}
\ee
contains the Doppler effect $\xi'(k)q$ with electron velocity $\xi'(k)$, the Edelstein effect $g'(k)q\sigma$ which couples Cooper pair momentum $q$ and electron spin $\sigma$ via SOC derivative $g'(k)$, and the Zeeman effect $2h\sigma$.
Near the superconducting phase transition, every Cooper pair with depairing energy $\mathcal{D}$ at temperature $T$ will increase the free energy by $\phi(x)\Delta$ on average, where $\Delta$ is the pairing potential, $\phi(x)={\rm Re}\{\psi(\frac{1}{2}+\frac{1}{2}ix)-\psi(\frac{1}{2})\}$ with $x=\mathcal{D}/(2\pi T)$ and $\psi(x)$ is the digamma function. 
We assume electrons with kinetic energy between $\xi\pm\Delta$ are involved in superconductivity, then the numbers of Cooper pairs near inner $(+)$ and outer $(-)$ Fermi points are $\rho_{\pm}\Delta$, with the corresponding density of states (DOS) $\rho_{\pm}$.
The overall free energy increase per pairing is hence
\be
\frac{\partial^2 f}{\partial\Delta^2}=
\sum_{\sigma=\pm}\rho_{\sigma}\phi\left(\frac{\mathcal{D}_{\sigma}}{2\pi T}\right)_{\rm F}+\rho\log\frac{T}{T_c},
\ee
with the total DOS $\rho=\rho_{+}+\rho_{-}$ and critical temperature $T_c$.
Here 2nd-order derivative $\partial^2f/\partial\Delta^2$ is taken at $\Delta=0$, and $\phi(x)_{\rm F}$ means the value of $\phi(x)$ at $k=k_{\rm F}$, with Fermi momentum defined by $\xi(k_{\rm F})=0$.

At weak fields near $T_c$, the free energy kernel reads
\be
K\equiv\frac{1}{\rho}\frac{\partial^2 f}{\partial\Delta^2}=t+(b_0-b_1q^2)qh+a_0q^2-a_1q^4
\ee
where $t=\log({T}/{T_c})=(T-T_c)/T_c$ is the reduced temperature, and the Ginzburg-Landau coefficients are
\bea\label{GLE_1D}
a_0=\frac{1}{4}\frac{C_0v_{\rm F}^2}{(\pi T_c)^2},\quad
b_0=\frac{C_0v_{\rm F}}{(\pi T_c)^2}(\delta+ \epsilon)\\\label{GLE_1Dc}
a_1=\frac{1}{8}\frac{C_1v_{\rm F}^4}{(\pi T_c)^4},\quad
b_1=\frac{C_1v_{\rm F}^3}{(\pi T_c)^4}(\delta+ 3\epsilon)
\eea
with numerical constants $C_0=7\zeta(3)/4=2.10$, $C_1=31\zeta(5)/32=1.00$ and two dimensionless quantities
\be\label{eq_SDEP}
\delta=\frac{\rho_{+}-\rho_{-}}{\rho_{+}+\rho_{-}},\quad
\epsilon=\frac{g'(k_{\rm F})}{\xi'(k_{\rm F})}.
\ee
Here, $\delta$ measures the DOS asymmetry between inner and outer Fermi points, and $\epsilon$ measures the Edelstein effect. Notice that $\epsilon$ does not directly measure the SOC strength but the SOC derivative, which is responsible for the Edelstein effect as elaborated in Eqs. (\ref{eq_v}) and (\ref{eq_dep}).

In 1D SDE, the magnitudes of critical currents $J_{c}^{\pm}$ along opposite directions can become unequal, quantified by the dimensionless supercurrent diode coefficient
\bea
\eta\equiv\frac{J_{c}^{+}-J_{c}^{-}}{J_{c}^{+}+J_{c}^{-}}\in [-1,1].
\eea

To be more specific, the field-boosted Cooper pair momentum $q_0$ and the diode coefficient $\eta$ can be obtained in terms of Ginzburg-Landau coefficients \cite{Yuan,Akito,James,Ilic}
\be
q_0=-\frac{b_0}{2a_0}h,\quad
\eta=\left(\frac{b_1}{b_0}-2\frac{a_1}{a_0}\right)\sqrt{\frac{|t|}{3a_0^3}}b_0h
\ee
which after some algebra turn out
\be\label{eq_1D}
q_0=-\frac{2h}{v_{\rm F}}(\delta+ \epsilon),\quad
\eta=1.21\frac{h}{h_P}|t|^{1/2}\epsilon
\ee
where $h_P=1.25T_c$ is the Pauli limiting field.
The field-boosted Cooper pair momentum is due to both DOS asymmetry and Edelstein effect $q_0\propto \delta+\epsilon$, while the field-induced SDE is due to the Edelstein effect only $\eta\propto\epsilon$.

When $\xi=k^2/(2m)-\mu$ and $g(k)=\alpha k$, it is found that $\delta\equiv 0$ and $\epsilon=\alpha/v_{\rm F}$ with $v_{\rm F}=\sqrt{2\mu/m}$, both $q_0$ and SDE are due to the Edelstein effect solely.

1D model Eq. (\ref{eq_1}) may describe quasi-1D superconductors such as transition metal trichalcogenides \cite{ZrTe3,HfTe3} and transition metal dichalcogenide nanotubes \cite{Qin}. 

In the following we will turn to more realistic 2D electron gas with SOC, and calculate SDE.

\textit{\textcolor{blue}{2D superconductors.}}---
In 2D, the critical current magnitude $J_{c}$ is a periodic function of its polar angle $\theta$,
\bea
J_{c}(\theta)=J_0\left\{1+\sum_{n=1}^{\infty}\eta_{n}\cos(n\theta+\phi_n)\right\},
\eea
where $\eta_n,\phi_n$ are Fourier parameters of the $n$-th order, which usually depend on temperature and field.
Inversion or time-reversal operation maps $J_c(\theta)$ to $J_c(\theta+\pi)$. Hence odd-$n$ terms $\eta_n,\phi_n$ describe the SDE, while even-$n$ terms $\eta_n,\phi_n$ describe the anisotropy of critical current.

Next we will calulate the leading order nonvanishing diode parameters $\eta_{2n+1},\phi_{2n+1}$ for the 2D superconductor with SOC $\bm g$ under an external Zeeman field $\bm h$,
\be
H=\xi({\bm k})+\bm g({\bm k})\cdot\bm\sigma+\bm h\cdot\bm\sigma,
\ee
where ${\bm k}=(k_x,k_y)$ is the 2D momentum vector, $\bm\sigma=(\sigma_x,\sigma_y,\sigma_z)$ denote Pauli matrices in spin space, and $\bm g,\bm h$ are vectors with three components. Due to time-reversal symmetry, $\xi(-\bm k)=\xi(\bm k)$ and $\bm g(-\bm k)=-\bm g(\bm k)$.

For generic $\bm g$ and $\bm h$, we find two bands $E_{\pm}=\xi\pm |\bm g+\bm h|$ in the normal phase, which correspond to two Fermi contours $E_{\pm}=0$, denoted as inner $(+)$ and outer $(-)$ Fermi contours as shown in Fig. \ref{fig1}(b).
At weak fields, the inter-pocket Cooper pairs formed by electrons from different Fermi contours are usually considered not favorable, and we consider intra-pocket Cooper pairs and examine the evolution of Fermi contours under external fields.
With the Rashba SOC $\bm g=\alpha_{\rm R}\bm k\times\hat{\bm z}$,
a finite in-plane Zeeman field shifts these two Fermi contours oppositely, resulting in Cartesian ovals without any rotation symmetry as elaborated in Ref. \cite{Yuan1}. 
The zeroth order critical current has full rotation symmetry, while the first order term has no rotation symmetry at all.
As a result, we expect the critical current $J_{c}(\theta)=J_0[1+\eta\cos(\theta+\phi)]$ up to the leading order.
With the Ising SOC $\bm g=g\hat{\bm z}$,
a finite out-of-plane Zeeman field deforms two Fermi contours into lima\c{c}cons with threefold rotation symmetry. 
We thus expect the critical current $J_{c}(\theta)=J_0[1+\eta\cos(3\theta+\phi)]$ up to the leading order.
Rashba and Ising SOCs can be realized in crystals with specific point groups. In the following, we will calculate the critical currents with specific point groups and corresponding SOCs.

The depairing energy of intra-pocket Cooper pairs is $\mathcal{D}_{\sigma}=E_{\sigma}\left(\frac{1}{2}\bm q+\bm k\right)-E_{\sigma}\left(\frac{1}{2}\bm q-\bm k\right)$ with $\sigma=\pm$ denoting Fermi contours type, similar to the 1D case in Eq. (\ref{eq_D1D}).
The free energy kernel can then be expressed as
\be
K=
\sum_{\sigma=\pm}\frac{\rho_{\sigma}}{\rho}\left\langle\phi\left(\frac{\mathcal{D}_{\sigma}}{2\pi T}\right)\right\rangle_{\rm F} +\log\frac{T}{T_c},
\ee
where $\rho_{\sigma}$ is the DOS of band $\sigma$, $\rho=\rho_{+}+\rho_{-}$ is the total DOS, and the Fermi contour average is
\be
\left\langle Q\right\rangle_{\rm F}\equiv\left.\oint_{\xi=0}Q(\bm k)dk\right/\oint_{\xi=0}dk.
\ee
Unlike 1D models, in 2D the Fermi contour average and hence free energy should be calculated numerically in most cases. 
In the rest of this paper we focus on special cases of SDE that can be analytically calculated.

\begin{table}[ht]
\centering
\caption{\textbf{The 1st order supercurrent diode effects.} Matrix $\Lambda$ in Eq. (\ref{eq_Lambda}) and phase $\phi$ in Eq. (\ref{eq_phi}) depend on the underlying point group $G$, and $\varphi$ is the polar angle of Zeeman field. For groups $C_n$ and $S_4$, $\varphi_0$ is not constraint by symmetries but determined by microscopic details. For groups $D_{n}$, $\varphi_0=0$ or $\pi$, for groups $C_{nv}$, $\varphi_0=\pm\pi/2$, and for group $D_{2d}$, $\varphi_0=0,\pi,\pm\pi/2$. In crystals, $n=3,4,6$.}
\begin{center}  
\begin{tabular}{c|c|c}  
\hline  $G$ & {$D_n,\ C_{nv},\ C_n$} & {$D_{2d},\ S_4$} \\
\hline
$\Lambda$ &{
$\begin{pmatrix}
    \cos\varphi_0 & \sin\varphi_0\\
    -\sin\varphi_0 & \cos\varphi_0
\end{pmatrix}$} & 
{
$\begin{pmatrix}
    \cos\varphi_0 & \sin\varphi_0\\
    \sin\varphi_0 & -\cos\varphi_0
\end{pmatrix}$}\\
\hline
$\phi$ &{$-\varphi+\varphi_0$} & {$\varphi+\varphi_0$} \\
\hline
\end{tabular}  
\end{center} 
\label{tab1}
\end{table}

When SOC vector is in-plane and linear in momentum
\be\label{eq_Lambda}
\bm g=\alpha\Lambda\bm k,\quad \Lambda\in {\rm O}(2),
\ee
with $\alpha>0$ and matrix $\Lambda$ under point groups in Table. \ref{tab1}, 
and the free energy kernel reads
\be
K=t+(b_0-b_1q^2)\bm q\Lambda^{-1}\bm h+a_0q^2-a_1q^4
\ee
where the Ginzburg-Landau coefficients are
\bea\label{eq_ising1}
a_0=\frac{1}{8}\frac{C_0v_{\rm F}^2}{(\pi T_c)^2},\quad
b_0=\frac{1}{2}\frac{C_0v_{\rm F}}{(\pi T_c)^2}(\delta+ \epsilon),\\\label{eq_ising2}
a_1=\frac{3}{64}\frac{C_1v_{\rm F}^4}{(\pi T_c)^4},\quad
b_1=\frac{3}{8}\frac{C_1v_{\rm F}^3}{(\pi T_c)^4}(\delta+ 3\epsilon).
\eea
with Fermi velocity $v_{\rm F}\equiv\sqrt{\left\langle |\nabla\xi|^2\right\rangle_{\rm F}}$, and the two dimensionless parameters $\delta$ and $\epsilon$ defined similar to Eq. (\ref{eq_SDEP}),
\be\label{eq_SDEP}
\delta=\frac{\rho_{+}-\rho_{-}}{\rho_{+}+\rho_{-}},\quad
\epsilon=\frac{\alpha}{v_{\rm F}}.
\ee
Compared with Eqs. (\ref{GLE_1D}, \ref{GLE_1Dc}) in 1D models, $a_0,b_0$ are multiplied by $\left\langle\cos^2\Theta\right\rangle_{\rm F}=1/2$ and $a_0,b_0$ by $\left\langle\cos^4\Theta\right\rangle_{\rm F}=3/8$, where $\Theta$ is the angle between $\bm k$ and $\bm q$.

As a result,
the field-boosted Cooper pair momentum is linear in field $\bm q_0=-2{\Lambda^{-1}\bm h}(\delta+ \epsilon)/{v_{\rm F}}$,
and the leading term of critical current Fourier series is the 1st order
\be\label{eq_phi}
\frac{J_c(\theta)}{J_0}=1+\eta\cos(\theta+\phi),\quad
\eta=1.28\frac{h_{\parallel}}{h_P}|t|^{1/2}\epsilon,
\ee
where $h_{\parallel}=\sqrt{h_x^2+h_y^2}$, and $\phi$ is determined by $\Lambda$ and the polar angle $\varphi$ of the Zeeman field as listed in Table. \ref{tab1}.

When $\xi=k^2/(2m)-\mu$, then $\delta=-\epsilon$. Cooper pair momentum is zero $\bm q_0\equiv\bm 0$, while SDE is finite. 

If the Edelstein effect is neglected $\epsilon=0$, the diode coefficient vanishes $\eta\equiv 0$ \cite{Ilic}, and further more, the in-plane upper critical field $H_{c2}$ becomes divergent at $T\to 0$ \cite{Dimi1,Dimi2,Samokhin}.
When Edelstein effect is included, the diode coefficient $\eta$ is nonzero, and $H_{c2}$ is finite at $T=0$ \cite{Yuan1,Fulde}.
Discussions of $H_{c2}$ will be given after the next section.

\textit{\textcolor{blue}{High-order supercurrent diode effect.}}---
By enumerating 2D crystal point groups without inversion, SDE could exist in 2D superconductors with point groups $D_n,C_{nv} $ $(n=3,4,6)$, $D_{2d},D_{3h}$ and their subgroups.
In the previous section, point groups $D_n,C_{nv},C_n (n=3,4,6)$ and $D_{2d},S_4$ have been studied, whose SDE is described by the 1st order harmonics of critical current as shown in Eq. (\ref{eq_phi}).
In the following we will consider SDE in point groups $D_{3h},C_{3h}$ and show that the leading nonzero harmonic of the critical current is the third order, due to nonlinear SOCs. Importantly, this high-order SDE is linear in external Zeeman fields. 

When the point group is $D_{3h}$, SOC is out-of-plane and cubic $\bm g=\beta{(k_x^3-3k_xk_y^2)\hat{\bm z}}$, the free energy kernel reads
\be
K=t +a_0q^2+b(q_x^3-3q_xq_y^2)h_z-a_1q^4
\ee
where $a_{0,1}$ are the same as Eqs. (\ref{eq_ising1}, \ref{eq_ising2}) and
\be
b=\frac{3}{4}\frac{C_1v_{\rm F}^3}{(\pi T_c)^4}\epsilon,\quad \epsilon=\frac{3\beta k_{\rm F}^2}{v_{\rm F}}.
\ee
The critical current Fourier series reads
\be
\frac{J_c(\theta)}{J_0}=1+\eta\cos 3\theta,\quad \eta=1.28\frac{h_z}{h_P}|t|^{1/2}\epsilon.
\ee
This threefold SDE $J_c(\theta+2\pi/3)=J_c(\theta)$ is purely due to the Edelstein effect of the out-of-plane Zeeman field, and may be realized in transition metal dichalcogenides such as NbSe$_2$ \cite{Lorenz} and twisted graphene systems where valley may be regarded as the pseudospin \cite{Yuan2,JXHu,LinJ,Diez}.
Notice that our previous discussion applies to $\Gamma$ pockets, and with multiple Fermi pockets at different valleys, $\epsilon$ should be understood as the averaged result over opposite valleys, which may cancel out to a small value \cite{Zhai}.

When the point group is $C_{3h}$, the critical current Fourier series is ${J_c(\theta)}/{J_0}=1+\eta\cos (3\theta+\phi)$ with an additional phase $\phi$ determined by microscopic details.

As the crystal rotations are limited to 2-, 3- 4- and 6-fold, and the Fourier series of SDEs have to be odd order, we conclude that SDEs can only be 1st or 3rd order, which are all listed in above.

To end our discussion on Edelstein effect and supercurrent diode effect, in the following section we would like to study the role of Edelstein effect in determining the in-plane upper critical field at low temperatures.

\textit{\textcolor{blue}{In-plane upper critical field.}}---
As the field increases, inter-pocket Cooper pairs will be favored, as elaborated in Fulde-Ferrell-Larkin-Ovchinnikov (FFLO) physics \cite{FF,LO}.
The depairing energy of inter-pocket Cooper pairs is $\mathcal{E}_{\sigma}=E_{\sigma}\left(\frac{1}{2}\bm q+\bm k\right)-E_{-\sigma}\left(\frac{1}{2}\bm q-\bm k\right)$ with $\sigma=\pm$ denoting Fermi contours type, and the total free energy kernel including both types of Cooper pairs is $K=\log({T}/{T_c})+K'$,
\be
K'=
\sum_{\sigma=\pm}\frac{\rho_{\sigma}}{\rho}\left\langle\phi\left(\frac{\mathcal{D}_{\sigma}}{2\pi T}\right)\sin^2\frac{\psi}{2}+\phi\left(\frac{\mathcal{E}_{\sigma}}{2\pi T}\right)\cos^2\frac{\psi}{2}\right\rangle_{\rm F}
\ee
where $\psi$ is the angle between $\bm h+\bm g(\frac{1}{2}\bm q\pm\bm k)$.
The in-plane upper critical field $H_{c2}(T)$ is then determined by
\be
\min_{\bm q}K(\bm q,T,\bm h)=0,
\ee
where $H_{c2}=|\bm h|/\mu_{\rm m}$ with electron magnetic moment $\mu_{\rm m}$, and $\bm h$ is in-plane. Here the orbital effect is neglected.

In superconductors with linear SOC $\bm g=\alpha\Lambda\bm k$, $H_{c2}(T)$ should depend on the dimensionless parameter $\delta$ and $\epsilon$. To get a clear physical picture, we consider two limits.

When DOS asymmtry is ignored, as calculated by Peter Fulde and his collaborators \cite{Fulde}, the in-plane upper critical field at zero temperature $H_{c2}^0\equiv H_{c2}(T=0)$ is always finite as constraint by Pauli and Edelstein effects,
\bea
\delta=0,\epsilon\neq 0:\quad H_{c2}^0=
    \frac{\Delta_0}{\mu_{\rm m}},\quad & \Delta_{so}<{\Delta_0}/{\sqrt{2}},\\
   H_{c2}^0=(2-\sqrt{2})\frac{\Delta_0+\Delta_{so}}{\mu_{\rm m}},\quad & \Delta_{so}>{\Delta_0}/{\sqrt{2}},
\eea
where $\Delta_0=1.76T_c$ is the zero-temperature pairing gap, $\Delta_{so}=\alpha k_{\rm F}$ is the SOC energy at Fermi momentum $k_{\rm F}$, and Cooper pair momentum is nonzero at $T=0$.


When Edelstein effect is neglected $\epsilon=0$, as shown in Refs. \cite{Dimi1,Dimi2,Samokhin}, $H_{c2}(T)$ diverges $H_{c2}(T)\propto T^{-\rho_{-}/\rho_{+}}$ as $T\to 0$,
where 
$\rho_{-}/\rho_{+}=(1-\delta)/(1+\delta)$ depends on $\delta$. This implies that DOS asymmetry alone is insufficient to kill superconductivity via Zeeman effect of in-plane fields.

When both DOS asymmetry and Edelstein effect are present, $H_{c2}(T)$ is in general finite at $T=0$,
\bea
\delta\epsilon\neq 0:\quad
H_{c2}^0=
    \frac{\Delta_0}{\mu_{\rm m}},\quad & \Delta_{so}<{\Delta_{so}^*},\\
   H_{c2}^0=\frac{\gamma_1\Delta_0+\gamma_2\Delta_{so}}{\mu_{\rm m}},\quad & \Delta_{so}>{\Delta_{so}^*},
\eea
where dimensionless parameters $\Delta_{so}^{*}/\Delta_0$ and $\gamma_{1,2}$ depend on $\delta$, which can be obtained numerically such as in Ref. \cite{Yuan1}. 
The Edelstein effect is significant in determining both $H_{c2}$ at high fields and SDE at low fields, and also plays an important role in magnetoelectric effects \cite{WYH1,WYH2}.

\textit{\textcolor{blue}{Conclusion.}}---
From microscopic models, we self-consistently work out the supercurrent diode effect under external Zeeman fields, which turns out to be directly related to the Edelstein effect. As a practical application, one may measure the Edelstein effect by the supercurrent diode effect, which maps out the microscopic information from the macroscopic measurement.


In monolayer superconductors, in-plane magnetic fields mainly induce Zeeman effect, and Edelstein effect is responsible for the SDE.
In multilayer superconductors, orbital effects \cite{LinJ,nature,Yuan3,Yuan2,Yuan1,WYH2} may also cause shift and deformation of Fermi contours, leading to the so-called orbital Edelstein effect, where valley or layer serves as the pseudospin. In these systems, SDE measurements would provide information on the orbital Edelstein effect.


\textit{\textcolor{blue}{Acknowledgement}}---
The author thanks K. T. Law for inspiring discussions.
This work is supported by the National Natural Science Foundation of China (Grant. No. 12174021).

\end{document}